\documentclass[
aps, 
pra,
superscriptaddress,
showpacs,
showkeys,
notitlepage,
twocolumn, 
numbers,
longbibliography]
{revtex4-1}

\usepackage[T1]{fontenc} 
\usepackage[utf8]{inputenc} 
\usepackage[english]{babel} 
\usepackage{graphicx, xcolor}
\usepackage{amsfonts, amssymb, amsmath, mathrsfs, calc, array, mathdesign, mathtools}
\usepackage[separate-uncertainty=true]{siunitx}
\usepackage[
bookmarksopen,
bookmarksdepth=2,
breaklinks=true
]{hyperref}
\usepackage[anythingbreaks]{breakurl}

\usepackage{abraces}
\renewcommand{\arraystretch}{0.45}

\usepackage{tabstackengine}
\stackMath
\newcolumntype{C}[1]{>{\centering\arraybackslash$}m{#1}<{$}}
   
\def\bra#1{\mathinner{\langle{#1}|}}
\def\ket#1{\mathinner{|{#1}\rangle}}
\def\braket#1{\mathinner{\langle{#1}\rangle}}
\def\Bra#1{\left\langle#1\right|}
\def\Ket#1{\left|#1\right\rangle}
\def\ontop#1#2{\setbox0\hbox{#2}\copy0\llap{\raise\ht0\hbox{#1}}}
\definecolor{darkblue}{rgb}{0,0,0.93} 
\definecolor{darkred}{rgb}{0.8,0,0} 
\hypersetup{
colorlinks=true, 
linkcolor=darkblue, 
citecolor=darkred, 
filecolor=darkblue, 
urlcolor=darkblue
}

\newcommand{\varxminus}{\beta^x_-}
\newcommand{\varxplus}{\beta^x_+}
\newcommand{\varyminus}{\beta^y_-}
\newcommand{\varyplus}{\beta^y_+}

\begin{document}

\title{Electromagnetic lattice gauge invariance in two-dimensional discrete-time quantum walks}

\author{Iv{\'a}n M{\'a}rquez-Mart{\'i}n}
\email{ivan.marquez@uv.es}
\affiliation{Departamento de F{\'i}sica Te{\'o}rica and IFIC, Universidad de Valencia and CSIC, Dr. Moliner 50, 46100 Burjassot, Spain}
\affiliation{Aix-Marseille Univ., Universit\'e de Toulon, CNRS, LIS, Marseille, France}

\author{Pablo Arnault}
\email{pablo.arnault@ific.uv.es}
\affiliation{Departamento de F{\'i}sica Te{\'o}rica and IFIC, Universidad de Valencia and CSIC, Dr. Moliner 50, 46100 Burjassot, Spain}

\author{Giuseppe Di Molfetta}
\email{giuseppe.dimolfetta@lis-lab.fr}
\affiliation{Aix-Marseille Univ., Universit\'e de Toulon, CNRS, LIS, Marseille, France}
\affiliation{Departamento de F{\'i}sica Te{\'o}rica and IFIC, Universidad de Valencia and CSIC, Dr. Moliner 50, 46100 Burjassot, Spain}

\author{Armando P{\'e}rez}
\email{armando.perez@uv.es}
\affiliation{Departamento de F{\'i}sica Te{\'o}rica and IFIC, Universidad de Valencia and CSIC, Dr. Moliner 50, 46100 Burjassot, Spain}

\begin{abstract}
Gauge invariance is one of the more important concepts in physics. We discuss this concept in connection with the unitary evolution of discrete-time quantum walks in one and two spatial dimensions, when they include the interaction with synthetic, external electromagnetic fields. One introduces this interaction as additional phases that play the role of gauge fields. Here, we present a way to incorporate those phases, which differs from previous works. Our proposal allows the discrete derivatives, that appear under a gauge transformation, to treat time and space on the same footing, in a way which is similar to standard lattice gauge theories. By considering two steps of the evolution, we define a density current which is gauge invariant and conserved. In the continuum limit, the dynamics of the particle, under a suitable choice of the parameters, becomes the Dirac equation, and the conserved current satisfies the corresponding conservation equation.
\end{abstract}

\keywords{Quantum walks, Quantum simulation, Lattice gauge theories}
\pacs{03.67.-a, 11.15.Ha, 03.75.-b, 47.11.Qr}

\maketitle

\section{Introduction}

Since its introduction in the electromagnetic theory, gauge invariance has been a paradigm in physics, and constitutes one of the main properties of successful theories such as the Standard Model of particle interactions \cite{book_Quigg13}.
On the one side, the gauge principle can be used as a guiding principle to define new theories, where the development of the Electroweak interaction theory is just an example. 
On the other side, the symmetry predicts the existence of a conserved current, which constitutes a powerful tool in the analysis of dynamical phenomena.

In this paper, we discuss the manifestation of U(1) gauge invariance within the context of a discrete-time quantum walk (DTQW) in a two-dimensional (2D) lattice, which could be generalized to 3D lattices. 
The dynamics of such DTQWs is driven by the action of unitary operators that act both on the spatial and internal degrees of freedom \cite{Meyer96a}. 
A particular interest in this gauge-invariant dynamical scheme arises from the possibility of describing with it, artificially, i.e.\ by engineering an appropriate spacetime dependence of the walker's phase, the effect of a magnetic field, or even a combination of electric and magnetic fields, on charged matter.
By itself, the magnetic field gives rise to interesting phenomena such as localization or controlled spreading \cite{Yalcinkaya2015} and Landau levels \cite{AD15}.
The magnetic field is also one of the main ingredients of the quantum Hall effect, with associated topological effects \cite{Kitagawa2012,Kitagawa2012b} and edge currents \cite{Verga2017}.
On the other hand, the combination of both a magnetic and an electric field exhibits richer features, like Bloch oscillations and the $\vec{E}\times\vec{B}$ drift  \cite{AD16}.
The observation of these effects with discrete-time schemes as we study here may be available in the future using internal-state- dependent transport of atoms in 2D optical lattices \cite{Groh2016, Brakhane16, SAMWA2018}, or of photons in 3D integrated-photonics circuits \cite{ONSY17}. In continuous-time schemes, atoms in optical lattices are also a promising platform \cite{JAKSCH200552, dalibard10a, Bloch2012, Dalibar2015} to observe such effects.

In order to consistently describe these effects with DTQWs, one needs to understand how U(1) gauge invariance can be incorporated within this framework, which differs notably from the electromagnetic theory in the continuum (i.e., in continuous spacetime). 
In fact, this is a general (serious) problem in physics, since going from the continuum to a lattice formulation is plagued with difficulties and new features \cite{book_montvay_munster_1994, Munster2000, book_smit_2002}. 
Moreover, the way of implementing gauge invariance in lattice models is usually not unique, with different approaches leading to the same limit in the continuum. 
Our proposal to achieve U(1) gauge invariance on the lattice exhibits close analogies both with the method used in quantum field theory \cite{Wilson74} and with recent works exhibiting similar but different U(1) lattice gauge invariances, in DTQWs \cite{DMD14, AD16, Arnault17} or in reversible cellular automata \cite{arrighi2018gauge}. We comment on the similarities and differences with these recent works.

This paper is organized as follows.
In Sec.\ \ref{sec:model}, we define a new family of DTQWs on a line, which satisfy a U(1) gauge invariance on the (1+1)D lattice. 
The discrete derivatives which intervene in this lattice gauge invariance treat time and space on the same footing, and are very much like those used in standard LGTs, in contrast with those of Refs.\ \cite{Arnault17,AD16}. 
This is achieved by applying the gauge-field exponentials either before or after the spatial shift, depending on whether the internal state of the walker is, say, up or down, respectively. 
We formally compute the continuum limit of these DTQWs, which concides, as desired and as in Refs.\ \cite{DMD14,Arnault17}, with the dynamics of a Dirac fermion in (1+1)D spacetime, coupled to a U(1), i.e., electro(magnetic) gauge field.  
In Sec.\ \ref{sec:model2D}, we extend the previous results to 2D walks, constructed by alternating 1D walks in the $x$ and $y$ directions of the spatial lattice. 
The way we ensure the U(1) lattice gauge invariance of this 2D scheme is by requiring it \emph{for each one-dimensional substep}, in contrast with the gauge invariance of Ref.\ \cite{AD16}. 
This ensures that time and space are still treated on the same footing at the level of the discrete derivatives, up to the fact that there are now, in 2D, two discrete derivatives in time, one for the even discrete-time coordinates, corresponding to the motion in the, say, $x$ direction and another one for the odd ones, corresponding to the motion in the $y$ direction.  
In Sec.\ \ref{sec:current}, finally, we derive analytically a lattice continuity equation, stating the conservation of a certain current on the lattice which is computed exactly. We comment on the differences between this continuity equation and that of Ref.\ \cite{AD16}.

\section{A new $\mathrm{U(1)}$ lattice gauge invariance for the DTQW on the line}\label{sec:model}

\subsection{Defining the 1D walk}

The state $\ket{\psi_{j}}$ of the walker at some arbitrary discrete time $j\in{\mathbb{N}}$, belongs to a Hilbert space $\mathcal{H}=\mathcal{H}_{\text{coin}} \otimes \mathcal{H}_{\text{position}}$. 
The Hilbert space $\mathcal{H}_{\text{position}}$ describes the external, spatial degree of freedom of the walker, and is spanned by the basis states $\{ \ket{x= p \epsilon} \}_{p \in \mathbb{Z}}$, where $\epsilon$ is the lattice spacing. 
The two-dimensional Hilbert space $\mathcal{H}_{\text{coin}}=\text{Span}\{\ket R,\ket L\}$ describes the internal, so-called coin degree of freedom of the walker, where `$R$' and `$L$' stand for `right' and `left'.
The projection of the walker's state on the position state $\ket{x= p \epsilon}$ at time $j$ is $\psi_{j,p} \equiv \langle x = p\epsilon | \psi_j\rangle$.
We identify $\ket R = (1,0)^{\top}$ and  $\ket L = (0,1)^{\top}$, where $\top$ denotes matrix transposition.
The dynamics of the DTQW is defined by its one- time-step evolution operator $U_j$, which is unitary and may depend on $j$,
\begin{equation}
\ket{\psi_{j+1}}=U_{j+1}\ket{\psi_{j}} \, .
\label{eq:UQW}
\end{equation}
As usual for DTQWs, the dynamics alternates between (i) rotations, $C$, of the coin degree of freedom, and (ii) spatial coin-state- dependent shifts, $S$:
\begin{equation}
U = S C \, ,
\label{eq:U_op}
\end{equation}
where, to lighten notations, the multiplication of $C$ by the identity tensor factor of the position Hilbert space has been, and will be, in similar cases, omitted.
We choose, for the coin rotation, the following one,
\begin{equation}
C(\theta) = e^{i\sigma^1\frac{\theta}{2}} = \begin{bmatrix}
\cos \frac{\theta}{2} & i \sin \frac{\theta}{2} \vspace{0.1cm} \\ i \sin \frac{\theta}{2} & \cos \frac{\theta}{2} \end{bmatrix} \, ,
\end{equation}
where $\sigma^n$ is the $n$th Pauli matrix, and $\theta$ is some angle, constant in time and uniform in position. 

Now, one of the novelties of the present work, is the way we gauge our walk. 
In Refs.\ \cite{DMD14,Arnault17,AD16}, gauging the walk amounts to gauge the standard coin-state- dependent shift, $S_{\text{free}} = e^{-i \sigma^3 \mathcal{K}}$, where $\mathcal{K}$ is the quasimomentum operator, as $S_{\text{free}} \rightarrow e^{i\alpha_{j}} S_{\text{free}} \, e^{-i \sigma^3 \xi_{j}}$, where $\alpha_{j,p}$ and $\xi_{j,p}$ are lattice counterparts of the temporal and spatial components of an electric potential of the continuum, $(A^0, A^1)$, with which they coincide in the continuum limit of the DTQW.
We have used the notation $\varphi_j : p \mapsto \varphi_{j,p}$ for diagonal operators in the position basis, such as $\alpha_j$ and $\xi_j$.
In the present work, we gauge the shift as follows: the relative order in which the shift and the gauge-field exponentials are applied, depend on the coin state, that is,
\begin{subequations}
\begin{align}
S({\alpha}_{j},{\xi}_{j}) &= \begin{bmatrix}
e^{-i \mathcal{K}} e^{i( \xi_{j} -\alpha_{j} )} & 0 \\
0 & e^{-i( \xi_{j} + \alpha_{j} )} e^{i \mathcal{K}}
\end{bmatrix}  \\
&= T e^{i (\beta_-)_{j}} \Lambda_{R} +e^{-i (\beta_+)_{j}}T^{\dagger} \Lambda_{L} \, ,
\end{align}
\end{subequations}
where $\dag$ denotes Hermitian conjugation.
We have introduced the following objects: (i) the translation operator by one lattice site to the right,
\begin{equation}
T = e^{-i\mathcal{K}} \, ,
\end{equation}
(ii) the two projectors associated to the coin space,
\begin{equation}
\Lambda_s = \ket s \! \! \bra s \, , \ \ \ \  s = R, L \, ,
\end{equation}
and (iii) the difference and sum of $\xi $ and $\alpha$,
\begin{subequations}
\begin{align}
\beta_- &= \xi -  \alpha  \\
\beta_+ &= \xi + \alpha \, .
\end{align}
\end{subequations}
The non-gauged coin-state- dependent shift is of course $S_{\text{free}} = S(0,0)$. 
We have chosen the superscripts $R$ and $L$ for, respectively, the upper and lower components of the wavefunction, because $S_{\text{free}}$ shifts the upper one to the right, and the lower one to the left.
To make notations clear, we introduce an auxiliary notation $\tilde{U}$ for the evolution operator, such that
\begin{subequations}
\begin{align}
U_{j} &\equiv \tilde{U}(\alpha_{j},\xi_{j},\theta) \\ 
&\equiv S(\alpha_{j},\xi_{j}) \, C(\theta)\, .
\end{align}
\end{subequations}

\subsection{Continuum limit of the 1D walk}

A first fact to mention is that this new way of gauging the walk does not change the continuum limit $\epsilon \rightarrow 0$.
Indeed, the fact that $e^{i\mathcal{K}}$ and $e^{i {f}(\mathcal{P})}$, where $\mathcal{P}$ is the position operator and $f$ an arbitrary function, do not commute, does make an important difference between the gauge procedure of the present work and that of Refs.\ \cite{DMD14,Arnault17,AD16} at the level of the DTQW, i.e.\ for a finite spacetime-lattice spacing. However, this becomes irrelevant in the continuum limit, since the latter is obtained by Taylor expanding all exponentials in their argument, and keeping only the first-order terms: in other words, at first order in their arguments, the exponentials always commute. 

Let us now recall this continuum limit $\epsilon \rightarrow 0$.  
Assume that, for a given quantity $Q$ defined on the spacetime lattice, $Q_{j,p}$ coincides with the value $Q{(t=j\epsilon,x=p\epsilon)}$ of some continuous function $Q$ of $t$ and $x$. 
First, rotate the coin state by a small amount at each time step, that is, set
\begin{equation}
\theta = - 2 \epsilon_{m} m \, ,
\end{equation}
with $\epsilon_m$ going to zero with $\epsilon$, which is the necessary condition for the continuum limit to exist; now, when going to the continuum, we will actually choose $\epsilon_m = \epsilon$, and the parameter $m$ will be identified as the mass of the walker.
Second, consider small gauge fields, that is, set
\begin{subequations}
\begin{align}
\alpha_{j,p} &= \epsilon_A  q A^0_{j,p} \\ 
\xi_{j,p} &= \epsilon_A  q A^1_{j,p} \, ,
\end{align}
\end{subequations}
with $\epsilon_A$ going to zero with $\epsilon$, which is also a necessary condition for the continuum limit to exist; again, when going to the continuum, we will actually choose $\epsilon_A = \epsilon$, and the parameter $q$ will be identified as the electromagnetic charge of the walker.
Assuming now that all $Q$'s are twice differentiable in both $t$ and $x$, and Taylor expanding the dynamics of the walker,  Eq.\ (\ref{eq:UQW}), at first order in $\epsilon$, delivers (i) zeroth-order terms that, by construction of our walk, cancel each other, which is a necessary condition for the continuum limit to exist, and (ii) first-order terms, which deliver a Hamiltonian equation that can be identified as the Dirac equation in (1+1)D spacetime, with a coupling to a U(1) (and thus Abelian) gauge field. 
This equation reads, in manifestly-covariant form,
\begin{equation}
\left(i\gamma^{\mu}_{\text{1D}} D_{\mu}-m\right)\psi=0 \, ,
\end{equation}
with $\mu=0,1$, the covariant derivative $D_{\mu}= \partial_{\mu} + iqA_{\mu}$, where
\begin{equation}
A_{0} =A^0 \, , \ \ \ \ A_{1} =- A^1 \, ,
\end{equation}
are the covariant components of the electric potential, and with the following gamma matrices,
\begin{equation}
\gamma^0_{\text{1D}} = \sigma^1 \, , \ \ \ \  \gamma^1_{\text{1D}} = -i\sigma^2  \, .
\end{equation}
As announced, we obtain, in the limit of small coin-rotation angles and small phases, the same continuum limit as if we had used the gauge procedure of Refs.\ \cite{DMD14,Arnault17,AD16}.

\subsection{A new $\mathrm{U(1)}$ lattice gauge invariance}

Our DTQW, Eq.\ (\ref{eq:UQW}), exhibits a remarkable U(1) lattice gauge invariance: it is invariant under local phase shifts of the form $\psi_{j,p} \rightarrow \psi'_{j,p}= e^{iq\chi_{j,p}} \psi_{j,p}$, where $\chi_{j,p}$ is an arbitrary space- and time-dependent quantity, provided the gauge fields become
\begin{align}
(A_{\mu}')_{j,p} &= (A_{\mu})_{j,p} - (d_{\mu} \chi)_{j,p} \, ,
\end{align}
for $\mu=0,1$, with
\begin{subequations} \label{eq:discrete_derivatives}
\begin{align} 
d_0 &= \frac{1}{\epsilon_A} \Delta_0 \Sigma_1 \\
d_1 &= \frac{1}{\epsilon_A} \Delta_1 \Sigma_0 \, ,
\end{align}
\end{subequations}
where  the $\Sigma$'s and $\Delta$'s act on sequences $Q_{j,p}$ of time and space as
\begin{subequations}
\begin{align}
(\Sigma_{\mu} Q)_{p_{\mu}} &= Q_{p_{\mu}+1} + Q_{p_{\mu}} \\
(\Delta_{\mu} Q)_{p_{\mu}} &= Q_{p_{\mu}+1} - Q_{p_{\mu}} \, ,
\end{align}
\end{subequations}
having introduced $p_0 \equiv j$ and $p_1 \equiv p$ for a more compact notation.
The discrete derivatives, Eqs.\ (\ref{eq:discrete_derivatives}), treat time and space on the same footing, on the contrary to those of Ref.\ \cite{AD16,Arnault17}. Morever, the $\Sigma$'s and $\Delta$'s defined here are sums and differences over one lattice spacing, or link between two sites, while in Ref.\ \cite{AD16,Arnault17} they were over two links. Notice that the $\Delta$'s are nothing but standard finite differences over one link. The fact that, here, one has to apply the $\Sigma$'s in addition to the $\Delta$'s, underlines that it may be appropriate that the gauge variables, that is, both the gauge fields and the local phase change, be defined on the links rather than on the sites, as in standard LGTs. We leave this matter to future work. Up to these extra $\Sigma$'s, the discrete derivatives involved in Eqs. (\ref{eq:discrete_derivatives}) are the same as those used in standard LGTs, that is, standard finite differences.

As done in Ref.\ \cite{Arnault17} for the 1D case, Ref.\ \cite{AD16} for the 2D case, and Ref.\ \cite{ADMDB16} for the non-Abelian 1D case, one can define a lattice counterpart  to the electromagnetic tensor in the continuum,
\begin{equation} \label{eq:electric_tensor}
(F_{\mu\nu})_{j,p} = (d_{\mu} A_{\nu})_{j,p}  - (d_{\nu} A_{\mu})_{j,p}  \, ,
\end{equation}
which is antisymmetric by construction. Since we are in 1D space, the only non-vanishing components are $(F_{01})_{j,p}=-(F_{10})_{j,p}$, which encode a lattice counterpart to the electric field, and there is no magnetic field. This quantity, $(F_{\mu\nu})_{j,p}$, is, as in the continuum, gauge-invariant by construction (on the spacetime lattice, obviously), since the $d_{\mu}$'s commute with each other.

In the continuum limit, $d_{\mu}$
tends towards the partial derivative $\partial_{\mu}$, the gauge transformation of Eq.\ (\ref{eq:discrete_derivatives}) becomes the standard one of the continuum, and the lattice counterpart to the electro(magnetic) tensor, Eq.\ (\ref{eq:electric_tensor}), becomes that electro(magnetic) tensor.

\section{2D generalization by alternating 1D walks along the $x$ and $y$ directions}\label{sec:model2D}

\subsection{Defining the 2D walk}

The walker can now move on a 2D lattice, and has spatial coordinates  $x=p\epsilon$ and $y=q\epsilon$, where $p,q \in \mathbb{Z}$.
 We will also use the notation $p=p_1$ and  $q=p_2$.
Now,  the 1D walk defined in the previous section admits a  2D generalization via a walk which alternates 1D walks in the $x$ and $y$ directions of the 2D lattice. This generalization reads
\begin{subequations} \label{eq:2D_main}
 \begin{align} 
\ket{\psi_{2l}}&=U^{(1)}_{2l} \ket{\psi_{2l-1}} \label{eq:2D_main_even}\\ 
\ket{\psi_{2l+1}}&=U^{(2)}_{2l+1} \ket{\psi_{2l}}\, ,
\label{eq:2D_main_odd}
\end{align}
\end{subequations}
with $l\in \mathbb{N}$ and where, for $i=1,2$,
\begin{subequations}
\begin{align}
U^{(i)}_{j} &\equiv \tilde{U}^{(i)}(\tfrac{1}{2} \alpha_{j}, \xi^i_{j},\theta^i) \\
&\equiv  S^{(i)}(\tfrac{1}{2} \alpha_{j}, \xi^i_{j}) \, C(\theta^i) \, ,
\end{align}
\end{subequations}
and
\begin{equation}
S^{(i)}(\tfrac{1}{2} \alpha_{j}, \xi^i_{j}) = T_{i} e^{i (\beta_-^{(i)})_{j}} \Lambda_{R} +e^{-i (\beta_+^{(i)})_{j}}T^{\dagger}_{i} \Lambda_{L} \, ,
\end{equation}
where
\begin{equation}
T_i = e^{-i \mathcal{K}_i} \, ,
\end{equation}
$\mathcal{K}_i$ being the quasimomentum operator along direction $i$, and
\begin{subequations}
\begin{align}
(\beta_-^{(i)})_{j,p,q} &= \xi^i_{j,p,q} - \frac{1}{2} \alpha_{j,p,q}  \\
(\beta_+^{(i)})_{j,p,q} &= \xi^i_{j,p,q}  + \frac{1}{2} \alpha_{j,p,q}  \, .
\end{align}
\end{subequations}

When the gauge fields, $\alpha_j$, $\xi^1_j$ and $\xi^2_j$, vanish, the alternate walk is translationally invariant in both time and space every two time steps. 
We will thus sometimes use the wording `substep' for the time evolutions $2l-1 \rightarrow 2l$ and $2l \rightarrow 2l+1$, and the wording `step' for $2l-1 \rightarrow 2l+1$.
We also introduce the two-substep walk,
\begin{align}
\ket{\psi_{2l+1}}=U^{\text{2D}}_{2l+1}\ket{\psi_{2l-1}} \, ,
\label{eq:2D_main_2}
\end{align}
where
\begin{align}
U^{\text{2D}}_{2l+1} = U^{(2)}_{2l+1} \, U^{(1)}_{2l} \, .
\end{align}

\subsection{Continuum limit for the 2D walk}

We perform the continuum limit of the two-substep walk.
Adapting the 1D-case procedure, we write
\begin{subequations}
\begin{align}
\alpha_{j,p,q} &= \epsilon_A  q A^0_{j,p,q} \\ 
\xi^{i}_{j,p,q} &= \epsilon_A  q A^i_{j,p,q} \, ,
\end{align}
\end{subequations}
for $i=1,2$.
Moreover, we choose
\begin{subequations}
\begin{align}
\theta^1 &= \frac{\pi}{2} - \epsilon_m m \\
\theta^2 &= - \frac{\pi}{2} - \epsilon_m m \, .
\end{align}
\end{subequations}
Assume now that, for a quantity defined on the spacetime lattice, $Q_{j,p,q}$ coincides with the value $Q(t=j\epsilon/2,x=p\epsilon,y=q \epsilon)$ of some continuous function $Q(t,x,y)$. The factor $1/2$ in the time variable is necessary to make the continuum limit of this two-substep DTQW match with the standard form of the Dirac equation.
Taking the continuum limit, $\epsilon\rightarrow 0$, of Eq.\ (\ref{eq:2D_main}), with $\epsilon_A = \epsilon_m = \epsilon$, we obtain
\begin{equation}
\left(i\gamma^{\mu}D_{\mu}-m\right)\psi=0 \, ,
\end{equation}
with
\begin{equation}
\gamma^0 = \sigma^1 \, , \ \ \ \gamma^1 = -i\sigma^3 \, , 
\ \ \ \gamma^2  = - i \sigma^2 \, .
\end{equation}

\subsection{Two-substep $\mathrm{U(1)}$ lattice gauge invariance}

By construction from 1D gauge-invariant walks, the 2D walk we have introduced,  Eq.\ (\ref{eq:2D_main}),  is invariant under the local phase shift $\psi_{j,p,q} \rightarrow \psi'_{j,p,q} = e^{iq\chi_{j,p,q}} \psi_{j,p,q}$, provided the gauge fields become
\begin{subequations}
\label{eq:2D_gauge_transfo}
\begin{align}
(A_{0}')_{j,p,q} &= \left\{  
\arraycolsep=1.4pt\def\arraystretch{1.5}
\begin{array}{ll}
(A_{0})_{j,p,q} - (d_{0}^1 \chi)_{j,p,q} & \ \ \ \text{for} \ j \ \text{even} \\
(A_{0})_{j,p,q} - (d_{0}^2 \chi)_{j,p,q} &\ \ \ \text{for} \ j \ \text{odd}
\end{array} 
 \right.
 \\ 
(A'_{i})_{j,p,q} &= (A_{i})_{j,p,q} - (d_{i} \chi)_{j,p,q} \, ,
\end{align}
\end{subequations}
for $i=1,2$, with
\begin{subequations} \label{eq:discrete_derivatives_2D}
\begin{align} 
d_0^k &= \frac{1}{\epsilon_A} \Delta_0 \Sigma_k \label{eq:discrete_derivatives_2D_time} \\
d_i &= \frac{1}{\epsilon_A} \Delta_i \Sigma_0 \, .
\end{align}
\end{subequations}
A first comment to make is that this 2D U(1) lattice gauge invariance differs from that of Ref.\ \cite{AD16} in the following: we have required, here, the gauge invariance \emph{for each one-dimensional substep}; we thus call this 2D U(1) lattice gauge invariance a two-substep gauge invariance.
In such a two-substep gauge invariance, $A_0$ transforms, by construction, differently at even and odd times: indeed, we have two different difference operators in time, $d^1_0$ for even times, and $d^2_0$ for odd times, which manifests the alternate construction of the walk.

Apart from this, the difference operators of Eq.\ (\ref{eq:discrete_derivatives_2D}) are a straightforward generalization of those used above in the 1D case, Eq.\ (\ref{eq:discrete_derivatives}). 
As in the 1D case, the difference operators of Eq.\ (\ref{eq:discrete_derivatives_2D}) treat space and time on the same footing (up to the two discrete derivatives in time), in constrast with Ref.\ \cite{AD16}. 
Additionally, in the present 2D case, these difference operators also treat the two directions of the lattice on the same footing, which is also in contrast with  Ref.\ \cite{AD16}.

Finally, one can define a lattice counterpart to the electromagnetic tensor, by generalizing the 1D lattice tensor, Eq.\ (\ref{eq:electric_tensor}), to the present 2D setting. Notice that one has to use the discrete temporal derivative $d^i_0$ in the definition of $F_{0i}$, so that $F_{01}$ and $F_{02}$ involve \emph{different} discrete temporal derivatives. This 2D lattice tensor has by construction the same properties of antisymmetry and of U(1) lattice gauge invariance as in the 1D case, and contains, additionally, a `lattice magnetic field' orthogonal to the 2D plane, namely, $(F_{12})_{j,p,q}$, for which there is no room in the 1D case.

In Ref.\ \cite{JWD18}, a three-substep U(1) lattice gauge invariance is suggested for a 2D DTQW on an equilateral triangular lattice, which is likely to be generalizable to the other DTQWs presented in this reference (isocele triangular and honeycomb lattices). 
There are two main differences between this work and the present one.
First, the correspondence between the spatial components of the lattice gauge field and those of the continuum is, in Ref.\ \cite{JWD18}, not one to one: three such components are needed on the lattice -- one for each substep --, while, the scheme being in 2D space, only two such components are needed in the continuum, which can be expressed as linear combinations of the three former ones, see Eqs.\ (18) of that reference.
In the present work, in contrast, the spatial components of the lattice gauge field match exactly those of the continuum.
This difference between Ref.\ \cite{JWD18} and the present work reflects the connectivity of the lattice, and calls for an understanding, in \emph{arbitrary} $n$D lattices, $n\in\mathbb{N}$, of the coupling of DTQWs to lattice counterparts of the electric and magnetic fields of the $n$D continuum.
Reference \cite{D18} opens the way to such an understanding.

The second main difference between Ref.\ \cite{JWD18} and the present work is that in the former, the relative order in which one applies the gauge field and the shift is not coin-state dependent, in contrast with ours.
As a consequence, the difference operators appearing in Ref.\ \cite{JWD18} do not treat time and space on the same footing as we do. 
More precisely, the temporal difference operator of Ref.\ \cite{JWD18}, see, e.g., the first equation of Eqs.\ (17) of that reference, is the same as that of the earlier work already mentioned previously, namely, Ref.\ \cite{AD16}\footnote{This is of course true up to the fact that only one `spatial symmetrization' $\sigma_i$, $i$ labelling the direction taken by the walker at each sub time step, appears in Ref.\ \cite{JWD18}, instead of the double one, $\Sigma_2\Sigma_1$, of the earlier reference \cite{AD16}, see the first equation of Eqs.\ (9) in that reference. This difference is simply due to the fact that in the earlier reference, the lattice gauge invariance is not multi-substep.}, and we already mentioned above that, in that earlier work, time and space are not treated on the same footing at the level of the difference operators, in contrast with the present work.

Eventually, notice the two following facts.
If one tries to impose to the 2D walk of Ref.\ \cite{AD16} a U(1) lattice gauge invariance for each one-dimensional substep, one needs at least to choose, for the corresponding gauge fields, linear combinations of those introduced in that reference -- for the no-substep gauge invariance --, but at \emph{different} spacetime-lattice sites.
The same thing happens when trying to impose, conversely, a no-substep U(1) lattice gauge invariance to the present 2D walk.

\section{Continuity equation and conserved current for the 2D walk}\label{sec:current}

In this section, we derive a lattice continuity equation from the dynamics of the DTQW, allowing us to introduce a current density which is both conserved and gauge invariant. 
In the whole section, we work on the spacetime lattice, and use the notations $t=j\epsilon$, $x=p\epsilon$ and $y=q\epsilon$, already introduced previously. 
By construction, the probability density at time $t$ and point $(x,y)$ is
\begin{equation}
J^{0}(t,x,y)=\bra{\psi_{t}}\Lambda_{x,y}\ket{\psi_{t}} \, ,
\end{equation}
where $\Lambda_{x,y}=\Ket{x,y} \! \! \Bra{x,y}$ is the projector on state $\Ket{x,y}$.

Now, another novelty of the present work with respect to Ref.\ \cite{AD16}, apart from the way we gauge our walk, discussed in the previous sections, is that we are going to derive our continuity equation and define the current density over \emph{two} time steps, i.e.\ $2\epsilon$, of Evolution (\ref{eq:2D_main_2}), i.e.\ four integers steps in the discrete-time variable $j$, since $t=j\epsilon/2$, while Ref.\ \cite{AD16} considers a single time step of this evolution to define the current density. As the reader shall see, this -- i.e.\ considering two time steps  to derive the continuity equation, instead of a single one -- will lead to the appearance of the standard (symmetric) finite difference as discrete derivatives, both in time and space, while the discrete derivatives of Ref.\ \cite{AD16} are more complicated, in particular the temporal one.

So, from this evolution over two time steps, one can easily derive a formula for the difference $J^{0}(t+\epsilon,x,y)-J^{0}(t-\epsilon,x,y)$, which can be written as 
\begin{align} \label{eq:differentesint}
&[\Delta^{\text{sym.}}_{0}J^{0}](t,x,y)= \\
& \ \ \ \ \frac{1}{2}\Bra{\psi_{t}}\left({U_{t+\epsilon}^{\text{2D}}}^{\dagger}\Lambda_{x,y}U_{t+\epsilon}^{\text{2D}}-U_{t-\epsilon}^{\text{2D}}\Lambda_{x,y}{U_{t-\epsilon}^{\text{2D}}}^{\dagger}\right)\Ket{\psi_{t}} \, , \nonumber
\end{align}
where, pay attention, we have used the following notation of the Hermitean conjugate for the backwards evolution, ${U^{\text{2D}}}^{\dagger}_{t-\epsilon} \equiv {U^{(1)}}^{\dagger}_{t-\epsilon/2} \, {U^{(2)}}^{\dagger}_{t}$, and where $\left[ \Delta_{0}^{\text{sym.}}f\right] (t,x,y)\equiv\frac{1}{2}\left[f(t+\epsilon,x,y)-f(t-\epsilon,x,y)\right]$, which defines a symmetric finite difference in time. 
We compute this quantity in App.\ \ref{sec:Continuity-equation}, and the result is given by Eq.\ (\ref{eq:continuityeq}). 
We can then recast Eq.\  (\ref{eq:continuityeq}), i.e. Eq.\ (\ref{eq:differentesint}), as 
\begin{equation} \label{eq:compactcontinuity}
\Delta_{\mu}^{\text{sym.}}J^{\mu}= 0 \, ,
\end{equation}
with implicit sum over $\mu=0,1,2$, and where we have introduced the symmetric finite differences in the $x$ and $y$ directions, $\left[\Delta_{1}^{\text{sym.}}f\right](t,x,y)\equiv\frac{1}{2}\left[f(t,x+\epsilon,y)-f(t,x-\epsilon,y)\right]$ and $\left[\Delta_{2}^{\text{sym.}}f\right](t,x,y)\equiv\frac{1}{2}\left[f(t,x,y+\epsilon)-f(t,x,y-\epsilon)\right]$. Equation (\ref{eq:compactcontinuity}) has the form of a continuity equation on the lattice. $J^{1} =J^{x}$ and $J^{2} = J^{y}$, appearing naturally as the current densities along the $x$ and $y$ directions, respectively, are defined by
\begin{align} \label{eq:Jx}
&J^{x}(t,x,y)=\bra{\psi_{t}} \Lambda_{x}\Big[ e^{i\varyminus(t,x,y)}e^{i\varyplus(t,x,y-\epsilon)}D_{2}M_{x}^{(1)}  \nonumber \\
& \ \ \ \ \ \ \  \ \ \ \ \ \ \ \ \ \ \ \ \  \ +e^{-i\varyplus(t,x,y-\epsilon)}e^{-i\varyminus(t,x,y)}D_{2}^{\dagger}M_{x}^{(2)} \nonumber \\
& \ \ \ \ \ \ \  \ \ \ \ \ \ \ \ \ \ \ \ \ \  +\Lambda_{y-\epsilon}M_{x}^{(3)} \nonumber \\
& \ \ \ \ \ \ \  \ \ \ \ \ \ \ \ \ \ \ \ \ \ +\Lambda_{y+\epsilon} M_{x}^{(4)}
\Big]\ket{\psi_{t}} \, ,
\end{align}
and
\begin{align} \label{eq:Jy}
&J^{y}(t,x,y)=\bra{\psi_{t}} \Lambda_{y}  \Big[ e^{i\varxplus(t+\frac{\epsilon}{2},x,y)} e^{i\varxminus(t+\frac{\epsilon}{2},x-\epsilon,y)}D_{1}M_{y}^{(1)}  \nonumber  \\
& \ \ \ \ \ \ \  \ \ \ \ \ \ \ \ \ \ \ \ \ \  +e^{-i\varxminus(t+\frac{\epsilon}{2},x-\epsilon,y)}e^{-i\varxplus(t+\frac{\epsilon}{2},x,y)}D_{1}^{\dagger}M_{y}^{(2)}  \nonumber \\
& \ \ \ \ \ \ \  \ \ \ \ \ \ \ \ \ \ \ \ \ \  +\Lambda_{x+\epsilon}M_{y}^{(3)}\nonumber \\
& \ \ \ \ \ \ \  \ \ \ \ \ \ \ \ \ \ \ \ \ \ +  \Lambda_{x-\epsilon}M_{y}^{(4)} \Big]\ket{\psi_{t}} \, ,
\end{align}
where we have used the following notations:
\begin{subequations}
\begin{align}
& \beta^{x}_{\pm} =\beta^{(1)}_{\pm} \, , \ \ \ \ \  \  \beta^{y}_{\pm} = \beta^{(2)}_{\pm} \, , \\
&S_x = S^{(1)} \, , \ \ \ \ \ \ S_y = S^{(2)} \, , \\
&C_x = C(\theta^1) \, , \ \ \ \, C_y = C(\theta^2) \, .
\end{align}
\end{subequations}
The rest of the notations we have introduced are defined in App.\ \ref{sec:Continuity-equation}.

Both the time and space differences are symmetric, which implies that they can be used to approximate true derivatives with a truncation error $O(\epsilon^{3})$, in contrast with the difference schemes over one time step, as that in Ref.\ \cite{AD16}, where the error is $O(\epsilon^{2})$.
There is a price to pay for this at the level of the discrete-spacetime scheme: the current is only defined at times $t$ which are even multiples of the time step $\Delta t$, while the walk is defined at all times -- less importantly, one needs in practice, in order to compute the current dynamics over a given area on a finite-size 2D lattice, more sites on the edges of that area with a two-step current than with a single-step one.

In terms of formal simplicity and connection to standard lattice gauge theories, notable advantages of the present continuity equation, Eq.\ (\ref{eq:compactcontinuity}), with respect to that of Ref.\ \cite{AD16}, is that the difference operators involved in it, i.e the $\Delta^{\text{sym.}}$'s, not only (i) treat all three spacetime coordinates on the same footing (while \emph{all} three are treated differently in Ref.\ \cite{AD16}), but (ii) correspond, in addition, to standard symmetric finite differences, while more complicated operators are used in Ref.\ \cite{AD16}.  
As in Ref.\ \cite{AD16}, however, the present difference operators intervening in the continuity equation are still different from those intervening in the gauge invariance.

It is easy to check (i) that the current densities defined above are gauge invariant under the transformations of Eq.\ (\ref{eq:2D_gauge_transfo}), and (ii) that Eq.\ (\ref{eq:compactcontinuity}) ensures the conservation of the total probability, i.e.\ $\sum_{x,y}J^{0}(t,x,y)$ does not change with time.

Eventually, notice the following. On the one hand, one can check that the present 2D DTQW, defined by Eqs. (\ref{eq:2D_main}), satisfies, in addition to the present two-step lattice continuity equation, a single-step one -- obtained by comparing the probability densities between two consecutive instants --, which has the same structure as that of Ref.\ \cite{AD16}, and involves, in particular, the same discrete derivatives -- the corresponding current is gauge invariant under the gauge transformations defined in the present work. On the other hand, one can also check that the 2D DTQW defined in Ref.\ \cite{AD16} satisfies, in addition to the single-step lattice continuity equation presented in that reference, a two-step one, which has the same structure as that of the present work, and involves, in particular, \emph{also symmetric finite differences as discrete derivatives} -- the corresponding current is gauge invariant under the gauge invariance of Ref.\ \cite{AD16}, which, recall, is different from the present one. These two combined results indicate that the `symmetrization' of the discrete derivatives when going from single-step to two-step continuity equations is independent from the way one gauges the walk, and is solely due to the alternate construction of the 2D walk. 

\section{Conclusion}

In this paper we have discussed some of the subtleties related to gauge invariance on discrete-time quantum walks that include the interaction with external, synthetic electromagnetic fields, appearing as additional phases related to those fields. As in standard lattice gauge theories, the way to introduce such interactions is not unique, and can lead to interesting new features. We introduce these additional phases in a way that differs from previous works in the literature. We have first described how this definition works for one-dimensional discrete-time quantum walks. This procedure has the advantage that the discrete derivatives which intervene in this lattice gauge invariance treat time and space on the same footing, and are very much like those used in standard LGTs, in contrast with those of Refs.\ \cite{Arnault17,AD16}. 

We extended the above dynamics to 2D lattices, by alternating 1D walks in the $x$ and $y$ directions of the spatial lattice, where we ensure the U(1) lattice gauge invariance of this 2D scheme by requiring it \emph{for each one-dimensional substep}, in contrast with the gauge invariance of Ref.\ \cite{AD16}. Also here, time and space are treated on the same footing at the level of the discrete derivatives -- up to the fact that there are now, in 2D space, two discrete derivatives in time, one for the even discrete-time coordinates, corresponding to the motion in the, say, $x$ direction, and another one for the odd ones, corresponding to the motion in the $y$ direction. 

By taking two time steps of the alternate walk, we introduced a density current which is both conserved and gauge invariant.
Both in the 1D and in the 2D cases, we have computed the continuum limit of these DTQWs. They coincide, as desired and as in Refs.\ \cite{DMD14,Arnault17} and \cite{Arnault17,AD16}, respectively, with the dynamics of a Dirac fermion in (1+1)D and (1+2)D spacetime, respectively, coupled to a U(1), i.e., electromagnetic gauge field. We also showed that, in two dimensions, the current conservation reproduces, in the continuum, that corresponding to the Dirac field. The procedure discussed here could be easily extended to the case of 3D lattices.

In our opinion, this work represents a sensible step on the way to quantum simulating the dynamics of a Dirac particle coupled to an external electromagnetic field. In addition to this, the quantum walk, as a dynamical process taking place on a lattice, introduces by itself new interesting phenomena, which are still to be fully explored even in the case of two dimensional lattices, which is the minimum dimensionality allowing for the description of both an electric and a magnetic field. 

Let us finally mention that a recent work \cite{CGWW2018} presents a unified framework to understand U(1) gauge invariance in discrete-time quantum walks on lattices and with coin spaces of arbitrary dimensions. This work should at the very least enlighten this field.

\begin{acknowledgments}

P.\ Arnault acknowledges inspiring discussions with Christopher Cedzich, Terry Farrelly and Reinhard F.\ Werner, and is grateful for their one-week hosting in the Quantum Information Group, Institut f\"ur Theoretische Physik, Leibniz Universit\"at Hannover, Germany.
This work has been funded by the ANR-12-BS02-007-01 TARMAC grant, 
the STICAmSud project 16STIC05 FoQCoSS, CSIC PIC2017FR6, the Spanish Ministerio
de Econom{\'i}a, Industria y Competitividad, MINECO-FEDER project FPA2017-84543-P,
SEV-2014-0398, and the Generalitat Valenciana grant GVPROMETEOII2014-087.
\end{acknowledgments}

\appendix

\section{\label{sec:Continuity-equation}Derivation of the continuity equation}

We start from Eq.\ (\ref{eq:differentesint}), with the purpose of obtaining the continuity equation, Eq.\ (\ref{eq:compactcontinuity}).
First, we work out the term
\begin{align}
U_{t-\epsilon}^{\text{2D}}\Lambda_{x,y}{U_{t-\epsilon}^{\text{2D}}}^{\dagger}  &\equiv   \left(T_{y}e^{i \varyminus(t,x,y)}M_{R_{y}}+e^{-i\varyplus(t,x,y)}T_{y}^{\dagger}M_{L_{y}}\right)\nonumber \\
 & \hspace{-1.3cm} \left(T_{x}e^{i \varxminus(t-\frac{\epsilon}{2},x,y)}M_{R_{x}}+e^{-i\varxplus(t-\frac{\epsilon}{2},x,y)}T_{x}^{\dagger}M_{L_{x}}\right)\Lambda_{x,y}\nonumber \\
 & \hspace{-1.3cm} \left(e^{-i \varxminus(t-\frac{\epsilon}{2},x,y)}T_{x}^{\dagger}M_{R_{x}}^{\dagger}+T_{x}e^{i\varxplus(t-\frac{\epsilon}{2},x,y)}M_{L_{x}}^{\dagger}\right)\nonumber \\
 & \hspace{-1.3cm} \left(e^{-i\varyminus(t,x,y)}T_{y}^{\dagger}M_{R_{y}}^{\dagger}+T_{y}e^{i\varyplus(t,x,y)}M_{L_{y}}^{\dagger}\right) \, ,
\end{align}
having used the notations
\begin{subequations}
\begin{gather}
T_{x} = T_{1} , \ \ \ T_{y} = T_{2}, \\
M_{s_i} = \Lambda_s C_i \, , \ \ \ s=R,L, \, \ \ \ i= x,y \, .
\end{gather}
\end{subequations}
After some tedious algebra, we arrive to 
\begin{align}
&U_{t-\epsilon}^{\text{2D}}\Lambda_{x,y}{U_{t-\epsilon}^{\text{2D}}}^{\dagger}= \Lambda_{x+\epsilon,y+\epsilon}M_{R_{y}}M_{R_{x}}M_{R_{x}}^{\dagger}M_{R_{y}}^{\dagger}\nonumber \\
 & \hspace{1.8cm}+e^{i\varyminus(t,x+\epsilon,y)}e^{i\varyplus(t,x+\epsilon,y-\epsilon)}\Lambda_{x+\epsilon}D_{2}M_{x}^{(1)}\nonumber \\
 & \hspace{1.8cm} +\Lambda_{x-\epsilon,y+\epsilon}M_{R_{y}}M_{L_{x}}M_{L_{x}}^{\dagger}M_{R_{y}}^{\dagger}\nonumber \\
 & \hspace{1.8cm} -e^{i\varyminus(t,x-\epsilon,y)}e^{i\varyplus(t,x-\epsilon,y-\epsilon)}\Lambda_{x-\epsilon}D_{2}M_{x}^{(1)}\nonumber \\
 & \hspace{1.8cm} +e^{-i\varyplus(t,x+\epsilon,y-\epsilon)}e^{-i\varyminus(t,x+\epsilon,y)}\Lambda_{x+\epsilon}D_{2}^{\dagger}M_{x}^{(2)}\nonumber \\
 & \hspace{1.8cm} +\Lambda_{x+\epsilon,y-\epsilon}M_{L_{y}}M_{R_{x}}M_{R_{x}}^{\dagger}M_{L_{y}}^{\dagger}\nonumber \\
 &  \hspace{1.8cm} -e^{-i\varyplus(t,x-\epsilon,y-\epsilon)}e^{-i\varyminus(t,x-\epsilon,y)}\Lambda_{x-\epsilon}D_{2}^{\dagger}M_{x}^{(2)}\nonumber \\
 & \hspace{1.8cm} +\Lambda_{x-\epsilon,y-\epsilon}M_{L_{y}}M_{L_{x}}M_{L_{x}}^{\dagger}M_{L_{y}}^{\dagger} \, .
\end{align}
Similarly, the Hermitian conjugate is given by 
\begin{align}
&{U_{t+\epsilon}^{\text{2D}}}^{\dagger}\Lambda_{x,y}{U_{t+\epsilon}^{\text{2D}}}= \Lambda_{x-\epsilon,y-\epsilon}M_{R_{x}}^{\dagger}M_{R_{y}}^{\dagger}M_{R_{y}}M_{R_{x}}\nonumber \\
 & \hspace{1.3cm} +e^{-i\varxminus(t+\frac{\epsilon}{2},x-\epsilon,y-\epsilon)}e^{-i\varxplus(t+\frac{\epsilon}{2},x,y-\epsilon)}D_{1}^{\dagger}\Lambda_{y-\epsilon}M_{y}^{(2)}\nonumber \\
 & \hspace{1.3cm}  +\Lambda_{x-\epsilon,y+\epsilon}M_{R_{x}}^{\dagger}M_{L_{y}}^{\dagger}M_{L_{y}}M_{R_{x}}\nonumber \\
 & \hspace{1.3cm} -e^{-i\varxminus(t+\frac{\epsilon}{2},x-\epsilon,y+\epsilon)}e^{-i\varxplus(t+\frac{\epsilon}{2},x,y+\epsilon)}D_{1}^{\dagger}\Lambda_{y+\epsilon}M_{y}^{(2)}\nonumber \\
 & \hspace{1.3cm} +e^{i\varxplus(t+\frac{\epsilon}{2},x,y-\epsilon)} 
 e^{i\varxminus(t+\frac{\epsilon}{2},x-\epsilon,y-\epsilon)}D_{1}\Lambda_{y-\epsilon}M_{y}^{(1)}\nonumber \\
 & \hspace{1.3cm} +\Lambda_{x+\epsilon,y-\epsilon}M_{L_{x}}^{\dagger}M_{R_{y}}^{\dagger}M_{R_{y}}M_{L_{x}}\nonumber \\
 & \hspace{1.3cm} -e^{i\varxminus(t+\frac{\epsilon}{2},x-\epsilon,y+\epsilon)}e^{i\varxplus(t+\frac{\epsilon}{2},x,y+\epsilon)}D_{1}\Lambda_{y+\epsilon}M_{y}^{(1)}\nonumber \\
 & \hspace{1.3cm} +\Lambda_{x+\epsilon,y+\epsilon}M_{L_{x}}^{\dagger}M_{L_{y}}^{\dagger}M_{L_{y}}M_{L_{x}} \, .
\end{align}
In the above equations, we have introduced the projectors
$\Lambda_{x}=\ket x \! \! \bra x$, $\Lambda_{y}=\ket y \! \! \bra y$, the operators $D_{1}=\Ket{x+\epsilon} \! \! \Bra{x-\epsilon}$,  $D_{2}=\Ket{y+\epsilon} \! \! \Bra{y-\epsilon}$, and
\begin{subequations}
\begin{align}
M_{x}^{(1)} &= M_{R_{y}}M_{R_{x}}M_{R_{x}}^{\dagger}M_{L_{y}}^{\dagger} \\
M_{x}^{(2)} &=  M_{L_{y}}M_{R_{x}}M_{R_{x}}^{\dagger}M_{R_{y}}^{\dagger} \\
M_{y}^{(1)} &=  M_{Lx}^{\dagger}M_{R_{y}}^{\dagger}M_{R_{y}}M_{Rx} \\
M_{y}^{(2)} &=  M_{R_{x}}^{\dagger}M_{R_{y}}^{\dagger}M_{R_{y}}M_{L_{x}} \, .
\end{align}
\end{subequations}
Performing in Eq.\ (\ref{eq:differentesint}) the substraction ${U_{t+\epsilon}^{\text{2D}}}^{\dagger}\Lambda_{x,y}U_{t+\epsilon}^{\text{2D}}-U_{t-\epsilon}^{\text{2D}}\Lambda_{x,y}{U_{t-\epsilon}^{\text{2D}}}^{\dagger}$ with the expressions obtained just above yields
\begin{widetext}
\begin{align}
2[\Delta^{\text{sym.}}_{0}J^{0}](t,x,y)&= \Lambda_{x+\epsilon,y+\epsilon}M_{\Lambda}^{(2)}+\Lambda_{x-\epsilon,y+\epsilon}M_{\Lambda}^{(4)}+\Lambda_{x+\epsilon,y-\epsilon}M_{\Lambda}^{(1)}+\Lambda_{x-\epsilon,y-\epsilon}M_{\Lambda}^{(3)}  \nonumber \\
&+\left(e^{i\varxplus(t+\frac{\epsilon}{2},x,y-\epsilon)}e^{i\varxminus(t+\frac{\epsilon}{2},x-\epsilon,y-\epsilon)}\Lambda_{y-\epsilon} 
 -e^{i\varxplus(t+\frac{\epsilon}{2},x,y+\epsilon)}e^{i\varxminus(t+\frac{\epsilon}{2},x-\epsilon,y+\epsilon)}\Lambda_{y+\epsilon}\right)D_{1}M_{y}^{(1)}  \nonumber \\ 
&+\left(e^{-i\varxminus(t+\frac{\epsilon}{2},x-\epsilon,y-\epsilon)}e^{-i\varxplus(t+\frac{\epsilon}{2},x,y-\epsilon)}\Lambda_{y-\epsilon}-e^{-i\varxminus(t+\frac{\epsilon}{2},x-\epsilon,y+\epsilon)}e^{-i\varxplus(t+\frac{\epsilon}{2},x,y+\epsilon)}\Lambda_{y+\epsilon}\right)D_{1}^{\dagger}M_{y}^{(2)}\nonumber \\
 & +\left(e^{i\varyminus(t,x-\epsilon,y)}e^{i\varyplus(t,x-\epsilon,y-\epsilon)}\Lambda_{x-\epsilon}-e^{i\varyminus(t,x+\epsilon,y)}e^{i\varyplus(t,x+\epsilon,y-\epsilon)}\Lambda_{x+\epsilon}\right)D_{2}M_{x}^{(1)}\nonumber \\
 & + \left(e^{-i\varyplus(t,x-\epsilon,y-\epsilon)}e^{-i\varyminus(t,x-\epsilon,y)}\Lambda_{x-\epsilon}-e^{-i\varyplus(t,x+\epsilon,y-\epsilon)}e^{-i\varyminus(t,x+\epsilon,y)}\Lambda_{x+\epsilon}\right)D_{2}^{\dagger}M_{x}^{(2)} \, ,\label{eq:continuityeq}
\end{align}
\end{widetext}
having introduced
\begin{subequations}
\begin{align}
M_{\Lambda}^{(1)} & =M_{L_{x}}^{\dagger}M_{R_{y}}^{\dagger}M_{R_{y}}M_{L_{x}}-M_{L_{y}}M_{R_{x}}M_{R_{x}}^{\dagger}M_{L_{y}}^{\text{\ensuremath{\dagger}}} \\
M_{\Lambda}^{(2)} & =M_{L_{x}}^{\dagger}M_{L_{y}}^{\dagger}M_{L_{y}}M_{L_{x}}-M_{R_{y}}M_{R_{x}}M_{R_{x}}^{\dagger}M_{R_{y}}^{\text{\ensuremath{\dagger}}} \\
M_{\Lambda}^{(3)} & =M_{R_{x}}^{\dagger}M_{R_{y}}^{\dagger}M_{R_{y}}M_{R_{x}}-M_{L_{y}}M_{L_{x}}M_{L_{x}}^{\dagger}M_{L_{y}}^{\text{\ensuremath{\dagger}}} \\
M_{\Lambda}^{(4)} & =M_{R_{x}}^{\dagger}M_{L_{y}}^{\dagger}M_{L_{y}}M_{R_{x}}-M_{R_{y}}M_{L_{x}}M_{L_{x}}^{\dagger}M_{R_{y}}^{\text{\ensuremath{\dagger}}} \! . \!
\end{align}
\end{subequations}
Now, one can check that the following relations hold,
\begin{subequations}
\begin{align}
M_{\Lambda}^{(1)} & =M_{y}^{(3)}-M_{x}^{(3)} \\
M_{\Lambda}^{(2)} & =-M_{y}^{(3)}-M_{x}^{(4)} \\
M_{\Lambda}^{(3)} & =M_{y}^{(4)}+M_{x}^{(3)} \\
M_{\Lambda}^{(4)} & =-M_{y}^{(4)}+M_{x}^{(4)} \, ,
\end{align}
\end{subequations}
having introduced
\begin{subequations}
\begin{align}
M_{x}^{(3)}&=\Lambda_{L}C_{y}\Lambda_{R}C_{y}^{\dagger}\Lambda_{L} \\
M_{x}^{(4)}&=\Lambda_{R}C_{y}\Lambda_{R}C_{y}^{\dagger}\Lambda_{R}-C_{x}^{\dagger}\Lambda_{L}C_{x} \\
M_{y}^{(3)}&=C_{x}^{\dagger}\Lambda_{L}C_{y}^{\dagger}\Lambda_{R}C_{y}\Lambda_{L}C_{x} \\
M_{y}^{(4)}&=C_{x}^{\dagger}\Lambda_{R}C_{y}^{\dagger}\Lambda_{R}C_{y}\Lambda_{R}C_{x}-\Lambda_L \, .
\end{align}
\end{subequations}
so that the above continuity equation, Eq.\ (\ref{eq:continuityeq}), can be recast as Eq.\ (\ref{eq:compactcontinuity}).

\section{Continuum limit of the current}

Let us check that the lattice continuity (or current-conservation) equation, Eq.\ (\ref{eq:continuityeq}), tends, in the continuum limit, towards the standard continuity equation involving the Dirac current $j^{\mu}=\bar{\psi}\gamma^{\mu}\psi$.
Taylor expanding the following quantities at first order in $\epsilon$ yields:
\begin{subequations}
\begin{align}
J_{0}(t+\epsilon,x,y)-J_{0}\left(t-\epsilon,x,y\right) & = 2\epsilon\partial_{t}J_{0}\left(t,x,y\right)  \\
\left(\Lambda_{x-\epsilon}-\Lambda_{x+\epsilon}\right) & =-2\epsilon\partial_{x}\Lambda_{x}  \\
\left(\Lambda_{y-\epsilon}-\Lambda_{y+\epsilon}\right) & =-2\epsilon\partial_{y}\Lambda_{y} \, .
\end{align}
\end{subequations}
Making use of Eqs.\ (\ref{eq:Jx}) and (\ref{eq:Jy}), one arrives to
\begin{align}
&\partial_{t}J_{0}\left(t,x,y\right) \\
& =-\partial_{x}\Bra{\psi_{t}}\Lambda_{xy}\Big(M_{x}^{(1)}+M_{x}^{(2)}+M_{x}^{(3)}+M_{x}^{(4)}\Big)\Ket{\psi_{t}}\nonumber \\
 & \ \ -\partial_{y}\Bra{\psi_{t}}\Lambda_{xy}\Big(M_{y}^{(1)}+M_{y}^{(2)}+M_{y}^{(3)}+M_{y}^{(4)}\Big)\Ket{\psi_{t}} \, . \nonumber
\end{align}
In our particular case, the coin matrices are, at zeroth order in $\epsilon$, 
$C_{x}=\frac{1}{\sqrt{2}}\begin{bmatrix}
1 & i\\
i & 1
\end{bmatrix}$ 
and 
$C_{y}=\frac{1}{\sqrt{2}}\begin{bmatrix}
1 & -i\\
-i & 1
\end{bmatrix}$, 
so that
\begin{subequations}
\begin{align}
M_{x}^{(1)}+M_{x}^{(2)}+M_{x}^{(3)}+M_{x}^{(4)} & =\begin{bmatrix}
0 & i\\
-i & 0
\end{bmatrix}=\gamma^{0}\gamma^{1}\\
M_{y}^{(1)}+M_{y}^{(2)}+M_{y}^{(3)}+M_{y}^{(4)}& =\begin{bmatrix}
1 & 0\\
0 & -1
\end{bmatrix}=\gamma^{0}\gamma^{2} \, .
\end{align}
\end{subequations}
Finally, the continuum limit of our continuity equation reads
\begin{align}
\partial_{t}J_{0}\left(t,x,y\right) & =-\partial_{x}\Bra{\psi_{t}}\Lambda_{xy}\gamma^{0}\gamma^{1}\Ket{\psi_{t}}\nonumber \\
 & -\partial_{y}\Bra{\psi_{t}}\Lambda_{xy}\gamma^{0}\gamma^{2}\Ket{\psi_{t}} \, .
\end{align}
This equation can be recast as $\partial_{\mu}(\bar{\psi}\gamma^{\mu}\psi)=0$,
the expected current-conservation equation.


\begin{thebibliography}{28}%
\makeatletter
\providecommand \@ifxundefined [1]{%
 \@ifx{#1\undefined}
}%
\providecommand \@ifnum [1]{%
 \ifnum #1\expandafter \@firstoftwo
 \else \expandafter \@secondoftwo
 \fi
}%
\providecommand \@ifx [1]{%
 \ifx #1\expandafter \@firstoftwo
 \else \expandafter \@secondoftwo
 \fi
}%
\providecommand \natexlab [1]{#1}%
\providecommand \enquote  [1]{``#1''}%
\providecommand \bibnamefont  [1]{#1}%
\providecommand \bibfnamefont [1]{#1}%
\providecommand \citenamefont [1]{#1}%
\providecommand \href@noop [0]{\@secondoftwo}%
\providecommand \href [0]{\begingroup \@sanitize@url \@href}%
\providecommand \@href[1]{\@@startlink{#1}\@@href}%
\providecommand \@@href[1]{\endgroup#1\@@endlink}%
\providecommand \@sanitize@url [0]{\catcode `\\12\catcode `\$12\catcode
  `\&12\catcode `\#12\catcode `\^12\catcode `\_12\catcode `\%12\relax}%
\providecommand \@@startlink[1]{}%
\providecommand \@@endlink[0]{}%
\providecommand \url  [0]{\begingroup\@sanitize@url \@url }%
\providecommand \@url [1]{\endgroup\@href {#1}{\urlprefix }}%
\providecommand \urlprefix  [0]{URL }%
\providecommand \Eprint [0]{\href }%
\providecommand \doibase [0]{http://dx.doi.org/}%
\providecommand \selectlanguage [0]{\@gobble}%
\providecommand \bibinfo  [0]{\@secondoftwo}%
\providecommand \bibfield  [0]{\@secondoftwo}%
\providecommand \translation [1]{[#1]}%
\providecommand \BibitemOpen [0]{}%
\providecommand \bibitemStop [0]{}%
\providecommand \bibitemNoStop [0]{.\EOS\space}%
\providecommand \EOS [0]{\spacefactor3000\relax}%
\providecommand \BibitemShut  [1]{\csname bibitem#1\endcsname}%
\let\auto@bib@innerbib\@empty
\bibitem [{\citenamefont {Quigg}(2013)}]{book_Quigg13}%
  \BibitemOpen
  \bibfield  {author} {\bibinfo {author} {\bibfnamefont {C.}~\bibnamefont
  {Quigg}},\ }\href {https://press.princeton.edu/titles/10156.html} {\emph
  {\bibinfo {title} {Gauge Theories of the Strong, Weak, and Electromagnetic
  Interactions}}},\ \bibinfo {edition} {2nd}\ ed.\ (\bibinfo  {publisher}
  {Princeton University Press},\ \bibinfo {year} {2013})\BibitemShut {NoStop}%
\bibitem [{\citenamefont {Meyer}(1996)}]{Meyer96a}%
  \BibitemOpen
  \bibfield  {author} {\bibinfo {author} {\bibfnamefont {D.~A.}\ \bibnamefont
  {Meyer}},\ }\bibfield  {title} {\bibinfo {title} {From quantum cellular
  automata to quantum lattice gases},\ }\href
  {https://link.springer.com/article/10.1007/BF02199356} {\bibfield  {journal}
  {\bibinfo  {journal} {{J}. {S}tat. {P}hys.}\ }\textbf {\bibinfo {volume}
  {85}},\ \bibinfo {pages} {551--574} (\bibinfo {year} {1996})}\BibitemShut
  {NoStop}%
\bibitem [{\citenamefont {Yal{\c{c}}{\i}nkaya}\ and\ \citenamefont
  {Gedik}(2015)}]{Yalcinkaya2015}%
  \BibitemOpen
  \bibfield  {author} {\bibinfo {author} {\bibfnamefont {{\.{I}}.}~\bibnamefont
  {Yal{\c{c}}{\i}nkaya}}\ and\ \bibinfo {author} {\bibfnamefont
  {Z.}~\bibnamefont {Gedik}},\ }\bibfield  {title} {\bibinfo {title}
  {Two-dimensional quantum walk under artificial magnetic field},\ }\href
  {\doibase 10.1103/physreva.92.042324} {\bibfield  {journal} {\bibinfo
  {journal} {Phys. Rev. A}\ }\textbf {\bibinfo {volume} {92}},\ \bibinfo
  {pages} {042324} (\bibinfo {year} {2015})}\BibitemShut {NoStop}%
\bibitem [{\citenamefont {Arnault}\ and\ \citenamefont
  {Debbasch}(2015)}]{AD15}%
  \BibitemOpen
  \bibfield  {author} {\bibinfo {author} {\bibfnamefont {P.}~\bibnamefont
  {Arnault}}\ and\ \bibinfo {author} {\bibfnamefont {F.}~\bibnamefont
  {Debbasch}},\ }\bibfield  {title} {\bibinfo {title} {Landau levels for
  discrete-time quantum walks in artificial magnetic fields},\ }\href
  {http://www.sciencedirect.com/science/article/pii/S0378437115006664}
  {\bibfield  {journal} {\bibinfo  {journal} {Physica A}\ }\textbf {\bibinfo
  {volume} {443}},\ \bibinfo {pages} {179--191} (\bibinfo {year}
  {2015})}\BibitemShut {NoStop}%
\bibitem [{\citenamefont {Kitagawa}(2012)}]{Kitagawa2012}%
  \BibitemOpen
  \bibfield  {author} {\bibinfo {author} {\bibfnamefont {T.}~\bibnamefont
  {Kitagawa}},\ }\bibfield  {title} {\bibinfo {title} {Topological phenomena in
  quantum walks: elementary introduction to the physics of topological
  phases},\ }\href {https://doi.org/10.1007/s11128-012-0425-4} {\bibfield
  {journal} {\bibinfo  {journal} {Quantum Inf. Process.}\ }\textbf {\bibinfo
  {volume} {11}},\ \bibinfo {pages} {1107--1148} (\bibinfo {year}
  {2012})}\BibitemShut {NoStop}%
\bibitem [{\citenamefont {Kitagawa}\ \emph {et~al.}(2012)\citenamefont
  {Kitagawa}, \citenamefont {Broome}, \citenamefont {Fedrizzi}, \citenamefont
  {Rudner}, \citenamefont {Berg}, \citenamefont {Kassal}, \citenamefont
  {Aspuru-Guzik}, \citenamefont {Demler},\ and\ \citenamefont
  {White}}]{Kitagawa2012b}%
  \BibitemOpen
  \bibfield  {author} {\bibinfo {author} {\bibfnamefont {T.}~\bibnamefont
  {Kitagawa}}, \bibinfo {author} {\bibfnamefont {M.~A.}\ \bibnamefont
  {Broome}}, \bibinfo {author} {\bibfnamefont {A.}~\bibnamefont {Fedrizzi}},
  \bibinfo {author} {\bibfnamefont {M.~S.}\ \bibnamefont {Rudner}}, \bibinfo
  {author} {\bibfnamefont {E.}~\bibnamefont {Berg}}, \bibinfo {author}
  {\bibfnamefont {I.}~\bibnamefont {Kassal}}, \bibinfo {author} {\bibfnamefont
  {A.}~\bibnamefont {Aspuru-Guzik}}, \bibinfo {author} {\bibfnamefont
  {E.}~\bibnamefont {Demler}}, \ and\ \bibinfo {author} {\bibfnamefont {A.~G.}\
  \bibnamefont {White}},\ }\bibfield  {title} {\bibinfo {title} {Observation of
  topologically protected bound states in photonic quantum walks},\ }\href
  {\doibase 10.1038/ncomms1872} {\bibfield  {journal} {\bibinfo  {journal}
  {Nat. Commun.}\ }\textbf {\bibinfo {volume} {3}},\ \bibinfo {pages} {882}
  (\bibinfo {year} {2012})}\BibitemShut {NoStop}%
\bibitem [{\citenamefont {Verga}(2017)}]{Verga2017}%
  \BibitemOpen
  \bibfield  {author} {\bibinfo {author} {\bibfnamefont {A.~D.}\ \bibnamefont
  {Verga}},\ }\bibfield  {title} {\bibinfo {title} {Edge states in a
  two-dimensional quantum walk with disorder},\ }\href
  {https://doi.org/10.1140/epjb/e2017-70433-1} {\bibfield  {journal} {\bibinfo
  {journal} {Eur. Phys. J. B}\ }\textbf {\bibinfo {volume} {90}},\ \bibinfo
  {pages} {41} (\bibinfo {year} {2017})}\BibitemShut {NoStop}%
\bibitem [{\citenamefont {Arnault}\ and\ \citenamefont
  {Debbasch}(2016)}]{AD16}%
  \BibitemOpen
  \bibfield  {author} {\bibinfo {author} {\bibfnamefont {P.}~\bibnamefont
  {Arnault}}\ and\ \bibinfo {author} {\bibfnamefont {F.}~\bibnamefont
  {Debbasch}},\ }\bibfield  {title} {\bibinfo {title} {Quantum walks and
  discrete gauge theories},\ }\href
  {https://link.aps.org/doi/10.1103/PhysRevA.93.052301} {\bibfield  {journal}
  {\bibinfo  {journal} {Phys. Rev. A}\ }\textbf {\bibinfo {volume} {93}},\
  \bibinfo {pages} {052301} (\bibinfo {year} {2016})}\BibitemShut {NoStop}%
\bibitem [{\citenamefont {Groh}\ \emph {et~al.}(2016)\citenamefont {Groh},
  \citenamefont {Brakhane}, \citenamefont {Alt}, \citenamefont {Meschede},
  \citenamefont {Asb\'oth},\ and\ \citenamefont {Alberti}}]{Groh2016}%
  \BibitemOpen
  \bibfield  {author} {\bibinfo {author} {\bibfnamefont {T.}~\bibnamefont
  {Groh}}, \bibinfo {author} {\bibfnamefont {S.}~\bibnamefont {Brakhane}},
  \bibinfo {author} {\bibfnamefont {W.}~\bibnamefont {Alt}}, \bibinfo {author}
  {\bibfnamefont {D.}~\bibnamefont {Meschede}}, \bibinfo {author}
  {\bibfnamefont {J.~K.}\ \bibnamefont {Asb\'oth}}, \ and\ \bibinfo {author}
  {\bibfnamefont {A.}~\bibnamefont {Alberti}},\ }\bibfield  {title} {\bibinfo
  {title} {Robustness of topologically protected edge states in quantum walk
  experiments with neutral atoms},\ }\href
  {https://link.aps.org/doi/10.1103/PhysRevA.94.013620} {\bibfield  {journal}
  {\bibinfo  {journal} {Phys. Rev. A}\ }\textbf {\bibinfo {volume} {94}},\
  \bibinfo {pages} {013620} (\bibinfo {year} {2016})}\BibitemShut {NoStop}%
\bibitem [{\citenamefont {Brakhane}(2016)}]{Brakhane16}%
  \BibitemOpen
  \bibfield  {author} {\bibinfo {author} {\bibfnamefont {S.}~\bibnamefont
  {Brakhane}},\ }\bibfield  {title} {\bibinfo {title} {The quantum walk
  microscope},\ }\href {http://hss.ulb.uni-bonn.de/2017/4609/4609.pdf}
  {\bibfield  {journal} {\bibinfo  {journal} {PhD thesis, Universit{\"a}t
  Bonn}\ } (\bibinfo {year} {2016})}\BibitemShut {NoStop}%
\bibitem [{\citenamefont {Sajid}\ \emph {et~al.}(2018)\citenamefont {Sajid},
  \citenamefont {Asb{\'o}th}, \citenamefont {Meschede}, \citenamefont
  {Werner},\ and\ \citenamefont {Alberti}}]{SAMWA2018}%
  \BibitemOpen
  \bibfield  {author} {\bibinfo {author} {\bibfnamefont {M.}~\bibnamefont
  {Sajid}}, \bibinfo {author} {\bibfnamefont {J.~K.}\ \bibnamefont
  {Asb{\'o}th}}, \bibinfo {author} {\bibfnamefont {D.}~\bibnamefont
  {Meschede}}, \bibinfo {author} {\bibfnamefont {R.~F.}\ \bibnamefont
  {Werner}}, \ and\ \bibinfo {author} {\bibfnamefont {A.}~\bibnamefont
  {Alberti}},\ }\bibfield  {title} {\bibinfo {title} {Creating {F}loquet
  {C}hern insulators with magnetic quantum walks},\ }\href
  {https://arxiv.org/abs/1808.08923} {\bibfield  {journal} {\bibinfo  {journal}
  {arXiv:1808.08923}\ } (\bibinfo {year} {2018})}\BibitemShut {NoStop}%
\bibitem [{\citenamefont {Boada}\ \emph {et~al.}(2017)\citenamefont {Boada},
  \citenamefont {Novo}, \citenamefont {Sciarrino},\ and\ \citenamefont
  {Omar}}]{ONSY17}%
  \BibitemOpen
  \bibfield  {author} {\bibinfo {author} {\bibfnamefont {O.}~\bibnamefont
  {Boada}}, \bibinfo {author} {\bibfnamefont {L.}~\bibnamefont {Novo}},
  \bibinfo {author} {\bibfnamefont {F.}~\bibnamefont {Sciarrino}}, \ and\
  \bibinfo {author} {\bibfnamefont {Y.}~\bibnamefont {Omar}},\ }\bibfield
  {title} {\bibinfo {title} {Quantum walks in synthetic gauge fields with
  three-dimensional integrated photonics},\ }\href
  {https://link.aps.org/doi/10.1103/PhysRevA.95.013830} {\bibfield  {journal}
  {\bibinfo  {journal} {Phys. Rev. A}\ }\textbf {\bibinfo {volume} {95}},\
  \bibinfo {pages} {013830} (\bibinfo {year} {2017})}\BibitemShut {NoStop}%
\bibitem [{\citenamefont {Jaksch}\ and\ \citenamefont
  {Zoller}(2005)}]{JAKSCH200552}%
  \BibitemOpen
  \bibfield  {author} {\bibinfo {author} {\bibfnamefont {D.}~\bibnamefont
  {Jaksch}}\ and\ \bibinfo {author} {\bibfnamefont {P.}~\bibnamefont
  {Zoller}},\ }\bibfield  {title} {\bibinfo {title} {The cold atom {H}ubbard
  toolbox},\ }\href
  {http://www.sciencedirect.com/science/article/pii/S0003491604001782}
  {\bibfield  {journal} {\bibinfo  {journal} {Ann. Physics}\ }\textbf {\bibinfo
  {volume} {315}},\ \bibinfo {pages} {52--79} (\bibinfo {year}
  {2005})}\BibitemShut {NoStop}%
\bibitem [{\citenamefont {Dalibard}\ \emph {et~al.}(2010)\citenamefont
  {Dalibard}, \citenamefont {Gerbier}, \citenamefont {Juzeli{\=u}nas},\ and\
  \citenamefont {{\"O}hberg}}]{dalibard10a}%
  \BibitemOpen
  \bibfield  {author} {\bibinfo {author} {\bibfnamefont {J.}~\bibnamefont
  {Dalibard}}, \bibinfo {author} {\bibfnamefont {F.}~\bibnamefont {Gerbier}},
  \bibinfo {author} {\bibfnamefont {G.}~\bibnamefont {Juzeli{\=u}nas}}, \ and\
  \bibinfo {author} {\bibfnamefont {P.}~\bibnamefont {{\"O}hberg}},\ }\bibfield
   {title} {\bibinfo {title} {Artificial gauge potential for neutral atoms},\
  }\href {https://link.aps.org/doi/10.1103/RevModPhys.83.1523} {\bibfield
  {journal} {\bibinfo  {journal} {{R}ev. Mod. Phys.}\ }\textbf {\bibinfo
  {volume} {83}},\ \bibinfo {pages} {1523} (\bibinfo {year}
  {2010})}\BibitemShut {NoStop}%
\bibitem [{\citenamefont {Bloch}\ \emph {et~al.}(2012)\citenamefont {Bloch},
  \citenamefont {Dalibard},\ and\ \citenamefont
  {Nascimb{\`{e}}ne}}]{Bloch2012}%
  \BibitemOpen
  \bibfield  {author} {\bibinfo {author} {\bibfnamefont {I.}~\bibnamefont
  {Bloch}}, \bibinfo {author} {\bibfnamefont {J.}~\bibnamefont {Dalibard}}, \
  and\ \bibinfo {author} {\bibfnamefont {S.}~\bibnamefont {Nascimb{\`{e}}ne}},\
  }\bibfield  {title} {\bibinfo {title} {Quantum simulations with ultracold
  quantum gases},\ }\href {\doibase 10.1038/nphys2259} {\bibfield  {journal}
  {\bibinfo  {journal} {Nat. Phys.}\ }\textbf {\bibinfo {volume} {8}},\
  \bibinfo {pages} {267--276} (\bibinfo {year} {2012})}\BibitemShut {NoStop}%
\bibitem [{\citenamefont {Dalibard}(2015)}]{Dalibar2015}%
  \BibitemOpen
  \bibfield  {author} {\bibinfo {author} {\bibfnamefont {J.}~\bibnamefont
  {Dalibard}},\ }\bibfield  {title} {\bibinfo {title} {Introduction to the
  physics of artificial gauge fields},\ }\href
  {https://arxiv.org/abs/1504.05520} {\bibfield  {journal} {\bibinfo  {journal}
  {arXiv:1504.05520}\ } (\bibinfo {year} {2015})}\BibitemShut {NoStop}%
\bibitem [{\citenamefont {Montvay}\ and\ \citenamefont
  {M{\"u}nster}(1994)}]{book_montvay_munster_1994}%
  \BibitemOpen
  \bibfield  {author} {\bibinfo {author} {\bibfnamefont {I.}~\bibnamefont
  {Montvay}}\ and\ \bibinfo {author} {\bibfnamefont {G.}~\bibnamefont
  {M{\"u}nster}},\ }\href {\doibase 10.1017/CBO9780511470783} {\emph {\bibinfo
  {title} {Quantum Fields on a Lattice}}},\ Cambridge Monographs on
  Mathematical Physics\ (\bibinfo  {publisher} {Cambridge University Press},\
  \bibinfo {year} {1994})\BibitemShut {NoStop}%
\bibitem [{\citenamefont {M\"{u}nster}\ and\ \citenamefont
  {Walzl}(2000)}]{Munster2000}%
  \BibitemOpen
  \bibfield  {author} {\bibinfo {author} {\bibfnamefont {G.}~\bibnamefont
  {M\"{u}nster}}\ and\ \bibinfo {author} {\bibfnamefont {M.}~\bibnamefont
  {Walzl}},\ }\bibfield  {title} {\bibinfo {title} {Lattice gauge theory - a
  short primer},\ }\href {https://arxiv.org/abs/hep-lat/0012005} {\bibfield
  {journal} {\bibinfo  {journal} {arXiv:hep-lat/0012005}\ } (\bibinfo {year}
  {2000})}\BibitemShut {NoStop}%
\bibitem [{\citenamefont {Smit}(2002)}]{book_smit_2002}%
  \BibitemOpen
  \bibfield  {author} {\bibinfo {author} {\bibfnamefont {J.}~\bibnamefont
  {Smit}},\ }\href {\doibase 10.1017/CBO9780511583971} {\emph {\bibinfo {title}
  {Introduction to Quantum Fields on a Lattice}}},\ Cambridge Lecture Notes in
  Physics\ (\bibinfo  {publisher} {Cambridge University Press},\ \bibinfo
  {year} {2002})\BibitemShut {NoStop}%
\bibitem [{\citenamefont {Wilson}(1974)}]{Wilson74}%
  \BibitemOpen
  \bibfield  {author} {\bibinfo {author} {\bibfnamefont {K.~G.}\ \bibnamefont
  {Wilson}},\ }\bibfield  {title} {\bibinfo {title} {Confinement of quarks},\
  }\href {https://link.aps.org/doi/10.1103/PhysRevD.10.2445} {\bibfield
  {journal} {\bibinfo  {journal} {Phys. Rev. D}\ }\textbf {\bibinfo {volume}
  {10}},\ \bibinfo {pages} {2445} (\bibinfo {year} {1974})}\BibitemShut
  {NoStop}%
\bibitem [{\citenamefont {{Di Molfetta}}\ \emph {et~al.}(2014)\citenamefont
  {{Di Molfetta}}, \citenamefont {Debbasch},\ and\ \citenamefont
  {Brachet}}]{DMD14}%
  \BibitemOpen
  \bibfield  {author} {\bibinfo {author} {\bibfnamefont {G.}~\bibnamefont {{Di
  Molfetta}}}, \bibinfo {author} {\bibfnamefont {F.}~\bibnamefont {Debbasch}},
  \ and\ \bibinfo {author} {\bibfnamefont {M.}~\bibnamefont {Brachet}},\
  }\bibfield  {title} {\bibinfo {title} {Quantum walks in artificial electric
  and gravitational fields},\ }\href
  {http://www.sciencedirect.com/science/article/pii/S0378437113011059}
  {\bibfield  {journal} {\bibinfo  {journal} {Physica A}\ }\textbf {\bibinfo
  {volume} {397}},\ \bibinfo {pages} {157--168} (\bibinfo {year}
  {2014})}\BibitemShut {NoStop}%
\bibitem [{\citenamefont {Arnault}(2017)}]{Arnault17}%
  \BibitemOpen
  \bibfield  {author} {\bibinfo {author} {\bibfnamefont {P.}~\bibnamefont
  {Arnault}},\ }\bibfield  {title} {\bibinfo {title} {Discrete-time quantum
  walk and gauge theories},\ }\href {https://arxiv.org/abs/1710.11123}
  {\bibfield  {journal} {\bibinfo  {journal} {PhD thesis, Universit{\'e} Pierre
  et Marie Curie}\ } (\bibinfo {year} {2017})}\BibitemShut {NoStop}%
\bibitem [{\citenamefont {Arrighi}\ \emph {et~al.}(2018)\citenamefont
  {Arrighi}, \citenamefont {Molfetta},\ and\ \citenamefont
  {Eon}}]{arrighi2018gauge}%
  \BibitemOpen
  \bibfield  {author} {\bibinfo {author} {\bibfnamefont {P.}~\bibnamefont
  {Arrighi}}, \bibinfo {author} {\bibfnamefont {G.~Di}\ \bibnamefont
  {Molfetta}}, \ and\ \bibinfo {author} {\bibfnamefont {N.}~\bibnamefont
  {Eon}},\ }\bibfield  {title} {\bibinfo {title} {A gauge-invariant reversible
  cellular automaton},\ }in\ \href
  {https://link.springer.com/chapter/10.1007/978-3-319-92675-9_1} {\emph
  {\bibinfo {booktitle} {Cellular Automata and Discrete Complex Systems}}}\
  (\bibinfo  {publisher} {Springer International Publishing},\ \bibinfo
  {address} {Cham},\ \bibinfo {year} {2018})\ pp.\ \bibinfo {pages}
  {1--12}\BibitemShut {NoStop}%
\bibitem [{\citenamefont {Arnault}\ \emph {et~al.}(2016)\citenamefont
  {Arnault}, \citenamefont {{Di Molfetta}}, \citenamefont {Brachet},\ and\
  \citenamefont {Debbasch}}]{ADMDB16}%
  \BibitemOpen
  \bibfield  {author} {\bibinfo {author} {\bibfnamefont {P.}~\bibnamefont
  {Arnault}}, \bibinfo {author} {\bibfnamefont {G.}~\bibnamefont {{Di
  Molfetta}}}, \bibinfo {author} {\bibfnamefont {M.}~\bibnamefont {Brachet}}, \
  and\ \bibinfo {author} {\bibfnamefont {F.}~\bibnamefont {Debbasch}},\
  }\bibfield  {title} {\bibinfo {title} {Quantum walks and non-{A}belian
  discrete gauge theory},\ }\href {\doibase 10.1103/physreva.94.012335}
  {\bibfield  {journal} {\bibinfo  {journal} {Phys. Rev. A}\ }\textbf {\bibinfo
  {volume} {94}},\ \bibinfo {pages} {012335} (\bibinfo {year}
  {2016})}\BibitemShut {NoStop}%
\bibitem [{\citenamefont {Jay}\ \emph {et~al.}(2018)\citenamefont {Jay},
  \citenamefont {Wang},\ and\ \citenamefont {Debbasch}}]{JWD18}%
  \BibitemOpen
  \bibfield  {author} {\bibinfo {author} {\bibfnamefont {G.}~\bibnamefont
  {Jay}}, \bibinfo {author} {\bibfnamefont {J.~B.}\ \bibnamefont {Wang}}, \
  and\ \bibinfo {author} {\bibfnamefont {F.}~\bibnamefont {Debbasch}},\
  }\bibfield  {title} {\bibinfo {title} {Dirac quantum walks on triangular and
  honeycomb lattices},\ }\href {https://arxiv.org/abs/1803.01304} {\bibfield
  {journal} {\bibinfo  {journal} {arXiv:1803.01304}\ } (\bibinfo {year}
  {2018})}\BibitemShut {NoStop}%
\bibitem [{\citenamefont {Debbasch}(2018)}]{D18}%
  \BibitemOpen
  \bibfield  {author} {\bibinfo {author} {\bibfnamefont {F.}~\bibnamefont
  {Debbasch}},\ }\bibfield  {title} {\bibinfo {title} {Action principles for
  quantum automata and {L}orentz invariance of discrete time quantum walks},\
  }\href {https://arxiv.org/abs/1806.02313} {\bibfield  {journal} {\bibinfo
  {journal} {arXiv:1806.02313}\ } (\bibinfo {year} {2018})}\BibitemShut
  {NoStop}%
\bibitem [{Note1()}]{Note1}%
  \BibitemOpen
  \bibinfo {note} {This is of course true up to the fact that only one `spatial
  symmetrization' $\sigma _i$, $i$ labelling the direction taken by the walker
  at each sub time step, appears in Ref.\ \cite {JWD18}, instead of the double
  one, $\Sigma _2\Sigma _1$, of the earlier reference \cite {AD16}, see the
  first equation of Eqs.\ (9) in that reference. This difference is simply due
  to the fact that in the earlier reference, the lattice gauge invariance is
  not multi-substep.}\BibitemShut {Stop}%
\bibitem [{\citenamefont {Cedzich}\ \emph {et~al.}(2018)\citenamefont
  {Cedzich}, \citenamefont {Geib}, \citenamefont {Werner},\ and\ \citenamefont
  {Werner}}]{CGWW2018}%
  \BibitemOpen
  \bibfield  {author} {\bibinfo {author} {\bibfnamefont {C.}~\bibnamefont
  {Cedzich}}, \bibinfo {author} {\bibfnamefont {T.}~\bibnamefont {Geib}},
  \bibinfo {author} {\bibfnamefont {A.~H.}\ \bibnamefont {Werner}}, \ and\
  \bibinfo {author} {\bibfnamefont {R.~F.}\ \bibnamefont {Werner}},\ }\bibfield
   {title} {\bibinfo {title} {Quantum walks in external gauge fields},\ }\href
  {https://arxiv.org/abs/1808.10850} {\bibfield  {journal} {\bibinfo  {journal}
  {arXiv:1808.10850}\ } (\bibinfo {year} {2018})}\BibitemShut {NoStop}%
\end{thebibliography}
\end{document}